\documentclass[10pt,twocolumn,aps,pra,showpacs]{revtex4}

\usepackage{bbm}

\begin{document}

\title{Postprocessing can speed up general quantum search algorithms}         
\author{Avatar Tulsi\\
        {\small Department of Physics, IIT Bombay, Mumbai-400076, India}}  

\email{tulsi9@gmail.com} 

\begin{abstract}

A general quantum search algorithm aims to evolve a quantum system from a known source state $|s\rangle$ to an unknown target state $|t\rangle$. It uses a diffusion 
operator $D_{s}$ having source state as one of its eigenstates and $I_{t}$, where $I_{\psi}$ denotes the selective phase inversion of $|\psi\rangle$ state. It evolves 
$|s\rangle$ to a 
particular state $|w\rangle$, call it w-state, in $O(B/\alpha)$ time steps where $\alpha$ is $|\langle t|s\rangle|$ and $B$ is a 
characteristic of the diffusion operator. Measuring the w-state gives the target state with the success probability of $O(1/B^{2})$ and $O(B^{2})$ applications of the algorithm 
can boost it from $O(1/B^{2})$ to $O(1)$, making the total time complexity $O(B^{3}/\alpha)$. In the special case of Grover's algorithm, $D_{s}$ is $I_{s}$ and $B$ is
very close to $1$. 
A more efficient way to boost the success probability is quantum 
amplitude amplification provided we can efficiently implement $I_{w}$. Such an efficient implementation is not known so far. In this paper, we 
present an efficient algorithm to approximate selective phase inversions of the unknown eigenstates of an operator using phase estimation algorithm. This algorithm is used
to efficiently approximate $I_{w}$ which reduces the time complexity of general algorithm to $O(B/\alpha)$. Though $O(B/\alpha)$ algorithms are known to exist, our algorithm 
offers physical implementation advantages. 

\end{abstract}

\pacs{03.67.Ac}

\maketitle

\section{INTRODUCTION}

Grover's algorithm or more generally quantum amplitude amplification drives a quantum computer from a \emph{source} state $|s\rangle$ to a \emph{target} state $|t\rangle$ by 
using selective phase inversion operators, $I_{s}$ and $I_{t}$, of these two states~\cite{grover,qaa1,qaa2}. The algorithm is an $O(1/\alpha)$
times iteration of the operator $\mathcal{A}(s,t) = -I_{s}I_{t}$ on $|s\rangle$ to get $|t\rangle$ where $\alpha = |\langle t|s\rangle|$. For search problem, we choose $|s\rangle$ 
to be the uniform superposition of all $N$ basis states to be searched i.e. $|s\rangle = \sum_{i}|i\rangle/\sqrt{N}$. In case of a unique solution, the target state 
$|t\rangle$ is a unique basis state and $\alpha = |\langle t|s\rangle| = 1/\sqrt{N}$. Thus Grover's algorithm outputs a solution in just $O(\sqrt{N})$ time steps which is 
quadratically faster than \emph{classical} search algorithms taking $O(N)$ time steps. Grover's algorithm is proved to be strictly optimal~\cite{optimal}.

A general framework of quantum search algorithms was presented in ~\cite{general}. The general quantum search algorithm replaces $I_{s}$ by a more general diffusion operator 
$D_{s}$ with the only restriction of having $|s\rangle$ as one of its eigenstates. This restriction seems to be more or less \emph{justified} as the 
diffusion operator should have some special connection with the source state. Let the normalized eigenspectrum of $D_{s}$ be given by 
$D_{s}|\ell\rangle = e^{\imath\theta_{\ell}}|\ell\rangle$ with $|\ell\rangle$ as the eigenstates and $e^{\imath\theta_{\ell}}$ ($\theta_{\ell}$) as the corresponding 
eigenvalues (eigenphases). Since a global phase is irrelevant in quantum dynamics, we 
choose $D_{s}|s\rangle = |s\rangle$, i.e. $\theta_{\ell=s} = 0$. The general search operator is $\mathcal{S} = D_{s}I_{t}$ and its iteration on $|s\rangle$ can be analyzed
by finding its eigenspectrum. Such an analysis was done in ~\cite{general} and we found that the performance of general algorithm depends upon two quantities 
$\Lambda_{1}$ and $\Lambda_{2}$, where
\begin{equation}
\Lambda_{p} = \sum_{\ell \neq s}|\langle \ell|t\rangle|^{2}\cot^{p}\frac{\theta_{\ell}}{2} \label{momentdefine}
\end{equation}
is the $p^{\rm th}$ moment of $\cot\frac{\theta_{\ell}}{2}$ with respect to the distribution $|\langle \ell |t\rangle|^{2}$ over all $\ell \neq s$. For a successful quantum 
search, we found $\Lambda_{1} = O(\alpha B) \approx 0$ to be an essential condition. Here $B=\sqrt{1+\Lambda_{2}}$ is a characteristic of the diffusion operator $D_{s}$. 
An algorithm was also presented in Section IV.A of ~\cite{general} which uses an 
ancilla qubit to cleverly control the applications of $D_{s}$ and $D_{s}^{\dagger}$ to effectively make $\Lambda_{1}$ zero. In this paper, we restrict ourselves 
to the case $\Lambda_{1} = 0$. In case, it is not so, we can always use just-mentioned algorithm to make it so. 

Unlike Grover's algorithm, the general algorithm does not rotate the quantum state in a plane spanned by $|s\rangle$ and $|t\rangle$. Rather, it induces rotation in a different
plane spanned by $|s\rangle$ and a particular state $|w\rangle$, which we refer as w-state here. The rotation angle is $2\alpha/B$ and w-state has an overlap of $O(1/B)$ 
with the desired final state $|t\rangle$. Thus, we get w-state after one round of 
algorithm which takes $O(B/\alpha)$ time steps and we perform $O(B^{2})$ rounds to get 
$|t\rangle$ with high probability. That makes the total time complexity to be $O(B/\alpha) \times O(B^{2}) = O(B^{3}/\alpha)$.

We know that performing $O(B^{2})$ rounds of algorithm to boost the success probability from $O(1/B^{2})$ to $O(1)$ is a classical process. Quantum mechanically, it can be
done more efficiently using quantum amplitude amplification (QAA). But to drive $|w\rangle$ towards $|t\rangle$, we need to implement $I_{w}$, the selective phase-inversion
of w-state. So far, $I_{w}$ is not known to be efficiently implementable. By definition $|w\rangle = \mathcal{S}^{q}|s\rangle$ for some $q$ so $I_{w} = \mathcal{S}^{q}I_{s}\mathcal{S}^{-q}$. So we can implement 
$I_{w}$ 
if we can implement $I_{s}$. But the whole purpose of studying general quantum search algorithm was to discuss the cases when $I_{s}$ is not efficiently implementable. Rather, 
what is available to us is a general diffusion operator $D_{s}$ of which $I_{s}$ is just a special case. We point out that $I_{s}$ is not easily implementable in cases of 
physical interest~\cite{spatial,kato,shenvi,realambainis}.

Thus, we are in some sense forced to use the classical process of running $O(B^{2})$ rounds of search algorithm to get the target state. However, in this paper, we 
show that $I_{w}$ can be efficiently implemented and $O(B/\alpha)$ time steps are enough to get the target state. We use 
quantum fourier transform to implement $I_{w}$. We point out that $O(B/\alpha)$ algorithms are known to exist~\cite{general,fastersearch} but our alternative approach
offers implementation advantages as in general, it uses significantly lesser number of controlled transformations which are physically harder to implement. In next section, 
we present the algorithm. The crux of our algorithm is the approximation of $I_{w}$ using phase estimation algorithm. We present the algorithm for this approximation in
Section III. We conclude in Section IV.

\section{ALGORITHM}

We briefly review the general quantum search algorithm~\cite{general}. It iterates the operator $D_{s}I_{t}^{\phi}$ on $|s\rangle$. Here $D_{s}$ is as defined earlier and 
$I_{t}^{\phi}$ is 
the selective phase rotation of target state by angle of $\phi$. We choose $\phi = \pi$ so that $I_{t}^{\phi}= I_{t}$. 
We assume $|s\rangle$ to be a non-degenerate eigenstate of $D_{s}$ for simplicity. 
The normalized eigenspectrum of $D_{s}$ is given by $D_{s}|\ell\rangle = e^{\imath\theta_{\ell}}|\ell\rangle$. By convention, $\theta_{\ell = s} = 0$. Let other 
eigenvalues satisfy
\begin{equation}
|\theta_{\ell \neq s}| \geq \theta_{\rm min} > 0,\ \ \theta_{\ell} \in [-\pi,\pi] \label{othereigenvalues}
\end{equation} 
The iteration of $\mathcal{S} =D_{s}I_{t}$ on $|s\rangle$ can be analysed by finding its eigenspectrum. The secular equation
was found in ~\cite{general} to be 
\begin{equation}
\sum_{\ell}|\langle \ell|t\rangle|^{2}\cot\frac{\lambda_{k}-\theta_{\ell}}{2} = 0. \label{secular}
\end{equation}
Any eigenvalue $e^{\imath \lambda_{k}}$ of $\mathcal{S}$ has to satisfy above equation.

Since $\cot x$ varies monotonically with $x$ except for the jump from $-\infty$ to $\infty$ when $x$ crosses zero, there is a unique 
solution $\lambda_{k}$ between each pair of consecutive $\theta_{\ell}$'s. As $\theta_{\ell = s} = 0$, there can be at most two solutions $\lambda_{k}$ in the 
interval $[-\theta_{\rm min},\theta_{\rm min}]$. Let these two solutions be $\lambda_{\pm}$. We have $|\lambda_{\pm}| < \theta_{\rm min}$. The eigenstates 
$|\lambda_{\pm}\rangle$ corresponding to these two eigenvalues $e^{\imath \lambda_{\pm}}$ are the only relevant eigenstates for our algorithm.
Precisely, if we assume $|\lambda_{\pm}| \ll \theta_{\rm min}$, then the initial state 
$|s\rangle$ is almost completely spanned by $|\lambda_{\pm}\rangle$. The eigenphases $\lambda_{\pm}$ are given by 
\begin{equation}
\lambda_{\pm} = \pm\frac{2\alpha}{B}(\tan \eta)^{\pm 1}\  ;\  \cot 2\eta = \frac{\Lambda_{1}}{2\alpha B}\ .\label{solutions}
\end{equation}
where
\begin{equation}
B = \sqrt{1 + \Lambda_{2}}\ ,\ \Lambda_{p} = \sum_{\ell \neq s}|\langle \ell|t\rangle|^{2}\cot^{p}\frac{\theta_{\ell}}{2}\ . \label{definitionLambda}
\end{equation}
We consider the case when $\Lambda_{1} = 0$. Then eq. (\ref{solutions}) indicates that 
\begin{equation}
\Lambda_{1} = 0 \Longrightarrow \eta = \frac{\pi}{4},\ \lambda_{\pm} = \pm\frac{2\alpha}{B}\ .
\end{equation} 

With $\eta = \pi/4$ and $\phi = \pi$, Eq. (23) and (24) of ~\cite{general} gives us the initial state $|s\rangle$ and the effect of iterating $\mathcal{S}$ on $|s\rangle$ in 
terms of two relevant eigenstates $|\lambda_{\pm}\rangle$. We have
\begin{equation}
|s\rangle = -\imath/\sqrt{2}[e^{\imath \lambda_{+}/2}|\lambda_{+}\rangle - e^{\imath \lambda_{-}/2}|\lambda_{-}\rangle], \label{slambdapmexpansion}
\end{equation}
and
\begin{equation}
\mathcal{S}^{q}|s\rangle =  -\imath/\sqrt{2} [e^{\imath q'\lambda_{+}}|\lambda_{+}\rangle - e^{\imath q'\lambda_{-}}|\lambda_{-}\rangle],
 \label{stateexpand}
\end{equation}
where $q'  = q+\frac{1}{2}$.

For $q = q_{\rm m} \approx \pi/2|\lambda_{\pm}| = \pi B/4\alpha$, the state $\mathcal{S}^{q_{\rm m}}|s\rangle$ is almost the w-state $|w\rangle$ given by 
\begin{equation}
\mathcal{S}^{q_{\rm m}}|s\rangle = |w\rangle = 1/\sqrt{2}(|\lambda_{+}\rangle + |\lambda_{-}\rangle). \label{wdefinedhere}
\end{equation}
As shown in ~\cite{general}, we have $|\langle t|w\rangle| = 1/B$. More explicitly,
\begin{equation}
|t\rangle = \frac{1}{B}|w\rangle + |\lambda_{\perp}\rangle, \label{targetintermsofeigenstates}
\end{equation}
where $|\lambda_{\perp}\rangle$ denotes any state orthogonal to $|\lambda_{\pm}\rangle$.
If we can efficiently implement $I_{w}$ then $O(B)$ iterations of quantum amplitude amplification
operator $\mathcal{A}(w,t) = I_{w}I_{t}$ on $|w\rangle$ will bring us close to the target state $|t\rangle$. 

We make a simple but useful observation that for our purpose $I_{w}$ is completely equivalent to $I_{\lambda_{\pm}}$, the selective phase inversion of the two-dimensional 
subspace
spanned by $|\lambda_{\pm}\rangle$. For our purpose, the quantum state is always of the form $\mathcal{A}^{q}(w,t)|w\rangle$ for an integer $q$.
As the amplitude amplification operator $\mathcal{A}(w,t)$ rotates the state vector in the plane spanned by $|w\rangle$ and 
$|t\rangle$, the state $\mathcal{A}^{q}(w,t)|w\rangle$ is always a linear combination of $|w\rangle$ and $|t\rangle$. Now, consider the two mutually orthogonal basis states 
$|\pm\rangle = |\lambda_{+}\rangle \pm |\lambda_{-}\rangle$ of $|\lambda_{\pm}\rangle$ subspace. We note that $|+\rangle = |w\rangle$ and Eq. (\ref{targetintermsofeigenstates}) 
implies that $|t\rangle$ also have zero component in $|-\rangle$ state. Thus $\mathcal{A}^{q}(w,t)|w\rangle$ has zero component in $|-\rangle$ state. For such a state, 
the operator $I_{\lambda_{\pm}}$, which is a selective phase inversion of both states $|+\rangle$ and $|-\rangle$, is completely equivalent to $I_{w}$, the selective phase
inversion of only $|+\rangle = |w\rangle$ state. Hence $I_{\lambda_{\pm}}$ is as good as $I_{w}$ for using quantum amplitude amplification to evolve $|w\rangle$ to $|t\rangle$.

The operator $I_{\lambda_{\pm}}$ is implemented using the phase estimation algorithm and our partial knowledge $|\lambda_{\pm}| < \theta_{\rm min}$ of the corresponding 
eigenphases. This knowledge helps us to distinguish $|\lambda_{\pm}\rangle$ from other 
eigenstates $|\lambda_{\perp}\rangle$ of $\mathcal{S}$. Next section shows that $O(\ln B/\theta_{\rm min})$ queries are sufficient to approximate $I_{\lambda_{\pm}}$ with an error of 
$o(1/B)$. Here $o(x) \ll x$ is the small-o notation used in computer science literature. As we need $O(B)$ applications of
$I_{w} = I_{\lambda_{\pm}}$, the total error in our algorithm is $Bo(1/B) \ll 1$ as errors
accumulate linearly in quantum mechanics. The total number of queries required to evolve $|w\rangle$ to $|t\rangle$ is
 $O(B) \times  O(\ln B/\theta_{\rm min}) = O(B\ln B/\theta_{\rm min})$. 

The total time complexity of our algorithm is the sum of (i) the time complexity $O(B/\alpha)$ of first evolution from $|s\rangle$ to $|w\rangle$ and (ii) 
the time complexity $O(B\ln B/\theta_{\rm min})$ of second evolution from $|w\rangle$ to $|t\rangle$. This is  
\begin{equation}
O\left(\frac{B}{\alpha} + \frac{B\ln B}{\theta_{\rm min}}\right).
\end{equation}
For typical cases when $\alpha \ll \theta_{\rm min}/\ln B$, above term is $O(B/\alpha)$. 

Our algorithm is not the first $O(B/\alpha)$ algorithm for general quantum searching 
but it offers implementation advantages over the earlier algorithms. To understand it,
we consider the controlled quantum search algorithm presented in ~\cite{general} having a time complexity of $O(B/\alpha)$. But that algorithm succeeds by using 
$O(B/\alpha)$ controlled applications of the operators $D_{s}$ and $I_{t}$. Similarly, recently a different $O(B/\alpha)$ algorithm was presented in ~\cite{fastergeneral}
but that also uses $O(B/\alpha)$ controlled applications of $D_{s}$ as well as $O(1/\alpha)$ controlled applications of $I_{t}$.
In comparison to these two algorithm, the alternative presented here uses only $O(B\ln B/\theta_{\rm min})$ controlled applications of $\mathcal{S} = D_{s}I_{t}$. These 
controlled applications are required for the Phase Estimation algorithm as we illustrate in the next section. For the typical cases, $\ln B/\theta_{\rm min} \ll 1/\alpha$ 
and hence the present algorithm uses
significantly lesser number of controlled operations compared to earlier algorithms. As controlled operations are harder to implement physically, our algorithm offers
implementation advantages. 

\section{Approximate Selective Phase Inversion}

\subsection{Basic Idea}

We present an algorithm to approximately implement the operator $I_{\lambda_{\pm}}$.  
We attach a workspace with the Hilbert space $\mathcal{H}_{w}$ to our main quantum system on which we wish to implement $I_{\lambda_{\pm}}$. We denote the Hilbert
space of main quantum system by $\mathcal{H}_{m}$ and refer to it as \emph{mainspace}. We work in the joint space $\mathcal{H}_{j} = \mathcal{H}_{m} \otimes \mathcal{H}_{w}$.
Let the initial state of this joint space be
\begin{equation}
|\xi,\sigma\rangle  = |\xi\rangle|\sigma\rangle, \label{xisigma}
\end{equation}
where $|\sigma\rangle$ is any standard known state of the workspace. We expand the initial state of mainspace $|\xi\rangle$ in the eigenbasis of the operator $\mathcal{S}$ as
\begin{equation}
|\xi\rangle = \sum_{\pm}\xi_{\pm}|\lambda_{\pm}\rangle + \sum_{h}\xi_{h}|\lambda_{\perp,h}\rangle, \label{xidefine}
\end{equation}
where $|\lambda_{\pm}\rangle$ are those two eigenstates on which we wish to implement the selective phase inversion and $|\lambda_{\perp,h}\rangle$ denote any of the 
remaining eigenstates orthogonal to $|\lambda_{\pm}\rangle$. The action of operator $I_{\lambda_{\pm}}$ on the mainspace is given by
\begin{equation}
\left(I_{\lambda_{\pm}} \otimes \mathbbm{1}_{w}\right)|\xi,\sigma\rangle = |\xi'\rangle|\sigma\rangle,
\end{equation}
where $\mathbbm{1}_{w}$ is the identity operator acting on the workspace and
\begin{equation}
|\xi'\rangle = I_{\lambda_{\pm}}|\xi\rangle = \sum_{\pm}\left(-\xi_{\pm}\right)|\lambda_{\pm}\rangle + \sum_{h}\xi_{h}|\lambda_{\perp,h}\rangle. \label{xiprimedefine}
\end{equation}
We point out that the index $h$ is not necessary in our analysis and we can represent all $|\lambda_{\perp,h}\rangle$ by $|\lambda_{\perp}\rangle$ for convenience. 

To approximate the operator $I_{\lambda_{\pm}}$ on the mainspace, we aim to obtain the state $|\xi'\rangle$. For this, we partition the workspace $\mathcal{H}_{w}$ into 
two mutually complementary subspaces: $\mathcal{Z}$ and $\mathcal{Z}^{\perp}$ which can be easily distinguished from each other. So we have 
$\mathcal{H}_{w} = \mathcal{Z} \cup \mathcal{Z}^{\perp}$.  Suppose we can design an operator $\mathcal{C}$ whose 
action on the joint space is given by
\begin{eqnarray}
\mathcal{C}\left(|\lambda_{\pm}\rangle|\sigma\rangle\right) &=& |\lambda_{\pm}\rangle|\mathcal{Z}\rangle + |O(\epsilon)\rangle, \nonumber \\ 
\mathcal{C}\left(|\lambda_{\perp}\rangle|\sigma\rangle\right) &=& |\lambda_{\perp}\rangle|\mathcal{Z}^{\perp}\rangle + |O(\epsilon)\rangle. \label{Cdefine}
\end{eqnarray} 
Here $|\mathcal{Z}\rangle$ and $|\mathcal{Z}^{\perp}\rangle$ denote a state completely within the subspaces $\mathcal{Z}$ and $\mathcal{Z}^{\perp}$ respectively. 
Also, $\epsilon \ll 1$ and $|O(\epsilon)\rangle$ denotes any arbitrary state of small length $O(\epsilon)$. Consider the following operator
\begin{equation}
\mathcal{R} = \mathcal{C}^{\dagger}\left(\mathbbm{1}_{m} \otimes I_{\mathcal{Z}}\right)\mathcal{C}  \label{Rdefine}
\end{equation}
where $I_{\mathcal{Z}}$ is the selective phase inversion of $\mathcal{Z}$-subspace acting on the workspace and 
$\mathbbm{1}_{m}$ is the identity operator acting on the mainspace. As $\mathcal{Z}$-subspace is easily distinguishable from its 
complementary subspace $\mathcal{Z}^{\perp}$, implementation of $I_{\mathcal{Z}}$ is straightforward.

Using Eq. (\ref{Cdefine}), we find that
\begin{eqnarray}
\left(\mathbbm{1}_{m} \otimes I_{\mathcal{Z}}\right)\mathcal{C}\left(|\lambda_{\pm}\rangle|\sigma\rangle\right) &=& -|\lambda_{\pm}\rangle|\mathcal{Z}\rangle + |O(\epsilon)\rangle, \nonumber \\ 
\left(\mathbbm{1}_{m} \otimes I_{\mathcal{Z}}\right)\mathcal{C}\left(|\lambda_{\perp}\rangle|\sigma\rangle\right) &=& |\lambda_{\perp}\rangle|\mathcal{Z}^{\perp}\rangle + |O(\epsilon)\rangle. \label{Cdefine2}
\label{2ndstage}
\end{eqnarray}
Also, left multiplication of Eq. (\ref{Cdefine}) by $\mathcal{C}^{\dagger}$ and the unitarity of $\mathcal{C}$ gives us
\begin{eqnarray}
\mathcal{C}^{\dagger}\left(|\lambda_{\pm}\rangle|\mathcal{Z}\rangle\right) & =& |\lambda_{\pm}\rangle|\sigma\rangle + |O(\epsilon)\rangle, \nonumber \\
\mathcal{C}^{\dagger}\left(|\lambda_{\perp}\rangle|\mathcal{Z}^{\perp}\rangle\right) & =& |\lambda_{\perp}\rangle|\sigma\rangle + |O(\epsilon)\rangle  
\label{Cdagger}
\end{eqnarray} 
Eqs. (\ref{Rdefine}), (\ref{2ndstage}) and (\ref{Cdagger}) imply that
\begin{eqnarray}
\mathcal{R}\left(|\lambda_{\pm}\rangle|\sigma\rangle\right) & = & -|\lambda_{\pm}\rangle|\sigma\rangle + |O(\epsilon)\rangle \nonumber \\
\mathcal{R}\left(|\lambda_{\perp}\rangle|\sigma\rangle\right) & = & |\lambda_{\perp}\rangle|\sigma\rangle + |O(\epsilon)\rangle  \label{Routput1} 
\end{eqnarray}
Using the linearity of $\mathcal{R}$ and Eqs. (\ref{xidefine},\ref{xiprimedefine}) along with the above equation, it is easy to check that
\begin{equation}
\mathcal{R}|\xi,\sigma\rangle = |\xi'\rangle|\sigma\rangle +|O(\epsilon)\rangle
\end{equation}
Thus we can approximate the operator $I_{\lambda_{\pm}}\otimes\mathbbm{1}_{w}$ and hence $I_{\lambda_{\pm}}$ upto an error of $O(\epsilon)$ provided we can design an
operator $\mathcal{C}$ satisfying Eq. (\ref{Cdefine}). The tolerable value of $\epsilon$ depends upon the total number of approximate 
implementations of $I_{\lambda_{\pm}}$ required by the algorithm. In our case, this number is $O(B)$ and $O(B)$ approximate implementations will contribute an error of 
$O(B\epsilon)$ as errors add up linearly. Thus for $\epsilon \ll 1/B$, our approximations 
will contribute a negligible error.

\subsection{Phase Estimation Algorithm}

Here we show that the phase estimation algorithm can be used to implement $\mathcal{C}$ satisfying Eq. (\ref{Cdefine}). Let the mainspace be in an eigenstate
$|\lambda\rangle$ of the operator $\mathcal{S}$ with the corresponding eigenvalue $e^{\imath \lambda}$. Here $|\lambda\rangle$ can be $|\lambda_{\pm}\rangle$ or  
$|\lambda_{\perp,h}\rangle$ for any $h$. We choose the workspace $\mathcal{H}_{w}$ to be the Hilbert space of
$\mu$ qubits. The initial state of workspace $|\sigma\rangle$ is chosen to be the state in which all $\mu$ qubits are in $|0\rangle$ state. We call this state all-zero state
and we denote it by $|0'\rangle$. Thus our initial state is $|\lambda,0'\rangle = |\lambda\rangle|0'\rangle$.

The operator $\mathcal{P}$ corresponding to the phase estimation algorithm is a successive application of three operators in the joint space, given by 
\begin{equation}
\mathcal{P} = (\mathbbm{1}_{N} \otimes \mathcal{F}^{\dagger})(c_{z}\mathcal{S}^{z})(\mathbbm{1}_{N} \otimes W)|\lambda,0'\rangle.
\end{equation} 
The first operator $\mathcal{P}_{1} = \mathbbm{1}_{N} \otimes W$ applies the Walsh-Hadamard transform on the workspace and leaves the mainspace unchanged. Walsh-Hadamard
transform is application of Hadamard gate on all $\mu$ qubits and it transforms $|0'\rangle$ state to a uniform superposition of all basis states $|z\rangle$ 
($z \in \{0,1,\ldots,2^{\mu}-1\}$) of the workspace. We have
\begin{equation}
\mathcal{P}_{1}|\lambda,0'\rangle = \frac{1}{2^{\mu/2}}\sum_{z = 0}^{2^{\mu}-1}|\lambda\rangle|z\rangle.
\end{equation}
The second operator $\mathcal{P}_{2} = c_{z}\mathcal{S}^{z}$ is a controlled application of $\mathcal{S}^{z}$ operator on the mainspace. It applies $z$ iterations of 
$\mathcal{S}$ on the mainspace if and only if the workspace is in the $|z\rangle$ state. We get 
\begin{equation}
\mathcal{P}_{2}\mathcal{P}_{1}|\lambda,0'\rangle = \frac{1}{2^{\mu/2}}\sum_{z = 0}^{2^{\mu}-1}(\mathcal{S}^{z}|\lambda\rangle)|z\rangle. \label{P2state1}
\end{equation}
As the mainspace is in an eigenstate $|\lambda\rangle$ of $\mathcal{S}$ and the corresponding eigenvalue is $e^{\imath \lambda}$, we get
\begin{equation}
\mathcal{P}_{2}\mathcal{P}_{1}|\lambda,0'\rangle = \frac{1}{2^{\mu/2}}\sum_{z = 0}^{2^{\mu}-1}e^{\imath \lambda z}|\lambda\rangle |z\rangle. \label{P2state}
\end{equation}

The third operator $\mathcal{P}_{3} = \mathbbm{1}_{N} \otimes \mathcal{F}^{\dagger}$ applies the inverse quantum fourier transform on the workspace but leaves the mainspace 
unchanged. The action of $\mathcal{F}^{\dagger}$ in a $2^{\mu}$-dimensional Hilbert space on each basis state $|z\rangle$ is given by
\begin{equation}
\mathcal{F}^{\dagger}|z\rangle = \frac{1}{2^{\mu/2}}\sum_{k =0}^{2^{\mu}-1}e^{-2\pi \imath kz/2^{\mu}} |k\rangle, \label{QFTequation}
\end{equation}
where $k$ is the numerical value of the binary number represented by the bit string encoded by the basis state $|k\rangle$. For example, for $|k\rangle = |0'\rangle$, 
all qubits are in $|0\rangle$ state and hence $k = 0$.
As suggested by Eq. (\ref{P2state}), before applying $\mathcal{F}^{\dagger}$, the workspace is in the state $(1/2^{\mu/2})\sum_{z}exp(\imath \lambda z)|z\rangle$. 
So, using (\ref{QFTequation}), we get
\begin{equation}
\mathcal{P}|\lambda,0'\rangle = \mathcal{P}_{3}\mathcal{P}_{2}\mathcal{P}_{1}|\lambda,0'\rangle = |\lambda\rangle|\phi_{\lambda}\rangle,  \label{mainworkstate}
\end{equation} 
where the state $|\phi_{\lambda}\rangle$ can be written using Eq. (\ref{QFTequation}) as 
\begin{equation}
|\phi_{\lambda}\rangle = \frac{1}{2^{\mu}}\sum_{k,z=0}^{2^{\mu}-1}exp\left[\imath\left(z\lambda-\frac{2\pi zk}{2^{\mu}}\right) \right]|k\rangle.  \label{phistatedefined}
\end{equation}
This is the standard output state of the phase estimation algorithm and has been analyzed well in literature (for example, see Sec. 5.2.1 of ~\cite{phase}). 
To get an idea, note that we have
\begin{equation}
\langle k|\phi_{\lambda}\rangle = \frac{1}{2^{\mu}}\sum_{z= 0}^{2^{\mu}-1}\left(exp\left[\lambda-\left(2\pi k/2^{\mu}\right)\right]\right)^{z}\ ,
\end{equation}
which is the sum of a geometric series, and after little calculation we get
\begin{equation}
|\langle k|\phi_{\lambda}\rangle| = \frac{1}{2^{\mu}} \frac{\sin(2^{\mu-1}\lambda - \pi k)}{\sin[(2^{\mu-1}\lambda - \pi k)/2^{\mu}]}. 
\end{equation}
What makes the state $|\phi_{\lambda}\rangle$ so useful is the fact that it provides a very good estimate of the eigenvalue $e^{\imath \lambda}$ of $\mathcal{S}$ 
provided the mainspace is in corresponding eigenstate $|\lambda\rangle$. The accuracy of this estimation increases with the increasing value of $\mu$, the number of
qubits used for phase estimation algorithm.

To elaborate it, we notice that each basis state $|k\rangle$ of the workspace corresponds to an integer number $k \in \{0,1,\ldots,2^{\mu}-1\}$ which is the numerical
value of the binary number encoded by the basis state. Let $\rm Prob(k)$ denote the probability of getting the basis state $|k\rangle$ after measuring the state
$|\phi_{\lambda}\rangle$. As shown in Sec. 5.2.1 of ~\cite{phase}, the probability distribution $\rm Prob(k)$ has a very sharp peak at 
$k = k_{\lambda}$, where
\begin{equation}
k_{\lambda} \stackrel{\rm nearest}{\approx}  2^{\mu}\frac{\lambda}{2\pi}\left(\rm mod 2^{\mu}\right)\  ,\label{numberdefine}
\end{equation}
where the symbol $\stackrel{\rm nearest}{\approx}$ denotes that L.H.S. is the nearest integer to the number on R.H.S. The probability $\rm Prob(k)$ falls very fast as we go 
away from its peak at $k = k_{\lambda}$.
To quantify this, let $c$ be a positive integer. We define a subspace $\mathcal{Y}_{\lambda}^{c}$ of the workspace $\mathcal{H}_{w}$ spanned by those basis
states $|k\rangle$ of $\mathcal{H}_{w}$ whose numerical value $k$ ranges from $(k_{\lambda} - c)(\rm mod 2^{\mu})$ to $(k_{\lambda} + c)(\rm mod 2^{\mu})$ 
in the increasing order.

For clarity, we give an example to elaborate on the increasing order. Suppose $2^{\mu} = 64$, $k_{\lambda} = 3$, $c = 5$. Then 
$(k_{\lambda} - c)(\rm mod 2^{\mu}) = (3-5)\rm mod 64 = 62$ and $(k_{\lambda} + c)(\rm mod 2^{\mu}) = (3+5)\rm mod 64 = 8$. By increasing order, we mean that
the basis states $|k\rangle$ of the subspace $\mathcal{Y}_{\lambda}^{c}$ for this example encode the numbers $\{62,63,64,1,2,\ldots,8\}$, not the other way around like 
$\{62,61,60,\ldots,9,8\}$ which is the decreasing order. On the other hand, if $k_{\lambda} = 62$ in this example, then 
$(k_{\lambda} - c)(\rm mod 2^{\mu}) = (62-5)\rm mod 64 = 57$ and $(k_{\lambda} + c)(\rm mod 2^{\mu}) = (62+5)\rm mod 64 = 3$ and 
the basis states $|k\rangle$ of $\mathcal{Y}_{\lambda}^{c}$ encode the numbers $\{57,58,\ldots,64,1,2,3\}$ not $\{57,56,\ldots,4,3\}$.

We now refer to Eq. (5.34) of Sec. 5.2.1 of ~\cite{phase}. It can be translated in our notation as
\begin{equation}
\rm Prob\left(k \in \mathcal{Y}_{\lambda}^{c}\right) = |\langle \mathcal{Y}_{\lambda}^{c}|\phi_{\lambda}\rangle|^{2} \geq 1-\frac{1}{2(c-1)} \stackrel{c \gg 1}{\approx} 1-\frac{1}{2c}\ . \label{probabilityeqn}
\end{equation}
where $|\langle \mathcal{Y}_{\lambda}^{c}|\phi_{\lambda}\rangle|$ denotes the magnitude of amplitude of $|\phi_{\lambda}\rangle$ state in $\mathcal{Y}_{\lambda}^{c}$ subspace
and $\rm Prob\left(k\in \mathcal{Y}_{\lambda}^{c}\right)$ denotes the probability of getting a basis state of $\mathcal{Y}_{\lambda}^{c}$ after measuring 
$|\phi_{\lambda}\rangle$ state. 

\subsection{Distinguishing eigenstates} 

To understand, how the phase estimation algorithm helps us to distinguish $|\lambda_{\pm}\rangle$ from $|\lambda_{\perp}\rangle$, we refer to the discussion after 
Eq. (\ref{secular}) at the beginning of this section. In our analysis of general quantum search algorithm, the assumption $|\lambda_{\pm}| \ll \theta_{\rm min}$
is crucial and we can write 
\begin{eqnarray}
|\lambda_{\pm}| &\leq& (1-2\delta)\theta_{\rm min}\ ,\ \ \ \delta \ll 1, \nonumber \\
|\lambda_{\perp}| &> &\theta_{\rm min}\ . \label{lambdabounds}
\end{eqnarray} 
Here the last inequality is due to the fact that $\lambda_{\pm}$ are the only eigenvalues of $\mathcal{S}$ in the interval 
$[-\theta_{\rm min}, \theta_{\rm min}]$.
We now define
\begin{equation}
\hat{\theta} = (1-\delta)\theta_{\rm min}\ ,\ \hat{k} \stackrel{\rm nearest}{\approx} 2^{\mu}\hat{\theta}/2\pi\ . \label{primedefine}
\end{equation}
With this definition and Eq. (\ref{lambdabounds}), it is easy to check the following equations.
\begin{eqnarray}
\theta \not\in \{-\hat{\theta},\hat{\theta}\} & \Longrightarrow & |\lambda_{\pm} - \theta| \geq \delta \theta_{\rm min}\ ,  \nonumber \\
\theta \in \{-\hat{\theta},\hat{\theta}\} & \Longrightarrow & |\lambda_{\perp} - \theta| \geq \delta \theta_{\rm min} \label{thetarelation12}
\end{eqnarray}

Let $\mathcal{X}$ be the subspace of $\mathcal{H}_{w}$ spanned by those basis states $|k\rangle$ whose numerical value $k$ ranges from $-\hat{k} (\rm mod 2^{\mu})$ to 
$\hat{k} (\rm mod 2^{\mu})$ in the increasing order. Let $\mathcal{X}^{\perp}$ be the complementary subspace of $\mathcal{X}$. Eqs. (\ref{numberdefine}), 
(\ref{primedefine}) and (\ref{thetarelation12}) imply that for all $|k\rangle \not\in \mathcal{X}$, $k$ differs from 
$k_{\lambda\pm}$ by at least the numerical value of 
\begin{equation}
\gamma = \frac{2^{\mu}\delta\theta_{\rm min}}{2\pi}
\end{equation}
By definition, such $|k\rangle$ are also the basis states of $\left(\mathcal{Y}_{\lambda \pm}^{\gamma}\right)^{\perp}$, which is the complementary subspace of 
$\mathcal{Y}_{\lambda\pm}^{\gamma}$. As $|k\rangle \not\in \mathcal{X} \Longrightarrow |k\rangle \not\in \mathcal{Y}_{\lambda\pm}^{\gamma}$, we have
$|k\rangle \in \mathcal{Y}_{\lambda\pm}^{\gamma} \Longrightarrow |k\rangle \in \mathcal{X}$ and hence $\mathcal{Y}_{\lambda\pm}^{\gamma}$ is a subspace of $\mathcal{X}$.
Using similar arguments, we can show that $\mathcal{Y}_{\lambda\perp}^{\gamma}$ is a subspace of $\mathcal{X}^{\perp}$. We write these facts as
\begin{equation}
\mathcal{Y}_{\lambda\pm}^{\gamma} \subseteq \mathcal{X}\ ,\ \mathcal{Y}_{\lambda\perp}^{\gamma} \subseteq \mathcal{X}^{\perp} \label{subsetrln}
\end{equation}
The value of $\gamma$ depends upon $\delta$ but the choice of $\delta$ is arbitrary as long as the condition $2\delta \ll 1$ is satisfied. We choose $\delta = 2\pi/128$ to get
\begin{equation}
\delta = 2\pi/128 \Longrightarrow \gamma = 2^{\mu-7}\theta_{\rm min}\ . \label{gammafix}
\end{equation} 

Suppose that the mainspace is in $|\lambda_{\pm}\rangle$ eigenstate of $\mathcal{S}$. Then Eqs. (\ref{probabilityeqn}), (\ref{subsetrln}) and (\ref{gammafix}) imply that the 
output state of workspace after phase estimation algorithm $|\phi_{\lambda\pm}\rangle$ is such that
\begin{equation}
|\langle \mathcal{X}|\phi_{\lambda\pm}\rangle|^{2} \geq |\langle \mathcal{Y}_{\lambda \pm}^{\gamma}|\phi_{\lambda\pm}\rangle|^{2} \geq 1-\frac{1}{2^{\mu-6}\theta_{\rm min}}\ , 
\label{componenteqn}
\end{equation}
where we have assumed $2^{\mu-6}\theta_{\rm min} \gg 1$. In above equations, $|\langle \mathcal{X}|\phi_{\lambda\pm}\rangle|$ is the component of $|\phi_{\lambda\pm}\rangle$ in 
$\mathcal{X}$-subspace. Using the fact $|\langle \mathcal{X}|\phi_{\lambda\pm}\rangle|^{2}+|\langle \mathcal{X}^{\perp}|\phi_{\lambda\pm}\rangle|^{2} = 1$ and above equation, 
we get
\begin{equation}
|\langle \mathcal{X}^{\perp}|\phi_{\lambda\pm}\rangle| \leq \sqrt{\frac{1}{2^{\mu-6}\theta_{\rm min}}} \ . \label{X1}
\end{equation}
Similar arguments show that if the mainspace is in $|\lambda_{\perp}\rangle$ state, we get the inequality
\begin{equation}
|\langle \mathcal{X}|\phi_{\lambda\perp}\rangle| \leq \sqrt{\frac{ 1}{2^{\mu-6}\theta_{\rm min}}} \ . \label{X2}
\end{equation}
In above two equations, we have not explicitly referred to the state of mainspace as that is implied in the definition of the workspace state
$(|\phi_{\lambda\pm}\rangle, |\phi_{\lambda\perp}\rangle)$ given by Eq. (\ref{mainworkstate}).
Assume that we know $\theta_{\rm min}$. Then we also know $\hat{\theta}$ and $\hat{k}$ and thus we can easily distinguish the subspace $\mathcal{X}$ from its complementary 
subspace $\mathcal{X}^{\perp}$. 

We choose $\mathcal{Z} = \mathcal{X}$ and $\mathcal{Z}^{\perp} = \mathcal{X}^{\perp}$
as then the criteria of easy distinguishability is satisfied. Then Eq. (\ref{Cdefine}) implies that the operator $\mathcal{P}$ corresponding to phase estimation algorithm can 
be chosen as the operator $\mathcal{C}$ with $|\sigma\rangle = |0'\rangle$. Using Eqs. (\ref{X1}) and (\ref{X2}), the corresponding $\epsilon$ is given by
\begin{equation}
\epsilon = O\left(\sqrt{\frac{1}{2^{\mu-6}\theta_{\rm min}}}\right). \label{epsilongamma}
\end{equation}
Thus the approximation error decreases and hence the accuracy increases as the number of ancilla qubits 
used for the phase estimation algorithm is increased. As we need $O(B)$ approximations, each approximation must have an error $ \epsilon = O(1/B)$. 
Then Eq. (\ref{epsilongamma}) implies that
\begin{equation}
\epsilon = O\left(\frac{1}{B}\right) \Longrightarrow 2^{\mu} = O\left(\frac{B^{2}}{\theta_{\rm min}}\right)\ . \label{naivescheme}
\end{equation}
Each application of the operator $\mathcal{P}$ or $\mathcal{P}^{\dagger}$ uses $2^{\mu}$ applications of the operator $\mathcal{S}$ to implement the operator
$c_{z}U^{z}$ as defined in Eq. (\ref{P2state1}). Hence the operator $\mathcal{R}$ (or $I_{\lambda_{\pm}}$) uses $2^{\mu+1}$ applications of $\mathcal{S}$ and equal number of 
oracle queries as each $\mathcal{S}$ requires an oracle query to implement $I_{t}$. We need $O(B)$ applications of $I_{\lambda_{\pm}}$ to transform $|w\rangle$ state to
the target state $|t\rangle$. With the value of $2^{\mu}$ given by Eq. (\ref{naivescheme}), the total number of queries used in our post-processing is
\begin{equation}
O\left(B \times \frac{B^{2}}{\theta_{\rm min}}\right) = O\left(\frac{B^{3}}{\theta_{\rm min}}\right)\ .
\end{equation} 
The number of ancilla qubits required by our algorithm is the logarithm of Eq. (\ref{naivescheme}), i.e.
\begin{equation}
\mu = 2\log_{2}B-\log_{2}\theta_{\rm min} + b, \label{mchoice1}
\end{equation}
where $b = O(1)$ is a small constant.

\subsection{Improving the query complexity}

We show that the query complexity of our post-processing algorithm can be improved by a factor of $O(B^{2}/\ln B)$ from $O(B^{3}/\theta_{\rm min})$ to $O(B\ln B/\theta_{\rm min})$.
We point out that in case of $B \not\gg 1$, we don't need any post-processing as then w-state has a 
significant overlap $O(1/B)$ with the target state. So for the cases of interest, $B \gg 1$ and hence improving the query complexity by a factor of $O(B^{2}/\ln B)$ is important.
To get this improvement, we don't choose $\mu$, the number of qubits for phase estimation algorithm, as given by Eq. (\ref{mchoice1}). Rather, we choose it to be
\begin{equation}
\mu = -\log_{2}\theta_{\rm min} + 16. \label{choicem}
\end{equation} 
Putting this choice in Eq. (\ref{epsilongamma}) gives us $\epsilon = O(2^{-5})$ which is a constant. If we stick to the phase estimation algorithm then the only way to further
reduce this error to its tolerable value $O(1/B)$ is to add more qubits for phase estimation. Eq. (\ref{epsilongamma}) implies that adding $2$ more qubits for phase estimation
will halve the value of $\epsilon$. Hence we add $2\log_{2}B$ qubits as in Eq. (\ref{mchoice1})  to make $\epsilon = O(1/B)$. Also, adding one qubit doubles the query 
complexity of phase estimation and hence the query complexity increases by a factor of $B^{2}$ just to reduce $\epsilon$ from $O(2^{-5})$ to $O(1/B)$. We show that once we 
achieve $\epsilon = O(2^{-5})$ using phase estimation, adding more qubits for phase estimation is not an efficient way to further reduce $\epsilon$. 
Rather, a better way is available which uses extra qubits in a more efficient way. 

We note that we are interested in only two mutually complementary and distinguishable subspaces $\mathcal{X}$ and $\mathcal{X}^{\perp}$ of the workspace. Let measurement of 
$|\phi_{\lambda}\rangle$ state output a state within these subspaces $\mathcal{X}$ and $\mathcal{X}^{\perp}$ with a probability of $\rm Prob_{\lambda}\left(\mathcal{X}\right)$ 
and 
$1-\rm Prob_{\lambda}\left(\mathcal{X}\right)$ 
respectively. With the choice (\ref{choicem}) of $\mu$ in Eqs. (\ref{X1}) and (\ref{X2}), we have 
\begin{eqnarray}
\rm Prob_{\lambda\pm}\left(\mathcal{X}^{\perp}\right) &=& \rm Prob_{\lambda\perp}\left(\mathcal{X}\right) = O(\epsilon^{2}) = O(2^{-10}) \nonumber \\
\rm Prob_{\lambda\pm}\left(\mathcal{X}\right) &=& \rm Prob_{\lambda\perp}\left(\mathcal{X}^{\perp}\right) = 1- O(2^{-10}). \label{probabilityexpressed}
\end{eqnarray} 
Suppose we have a single qubit in a state $|\beta_{\lambda}\rangle$ which simulates this measurement statistics which means that measuring $|\beta_{\lambda}\rangle$ state 
gives the output $1$ with the probability $\rm Prob_{\lambda}\left(\mathcal{X}\right)$ and $0$ with the probability $1-\rm Prob_{\lambda}\left(\mathcal{X}\right)$. We write it 
as
\begin{equation}
|\beta_{\lambda}^{\pm}\rangle = \cos \beta_{\lambda} |0\rangle \pm \imath \sin \beta_{\lambda}|1\rangle,\ \ \sin \beta_{\lambda} = |\sqrt{\rm Prob_{\lambda}\left(\mathcal{X}\right)}|.  
\label{betadefine} 
\end{equation} 
We could have chosen any phase factor for the $|1\rangle$ state in above equation but the choice $\pm$ is sufficient for our purpose. The above single qubit state is a very 
important resource to improve our approximation. To understand this, suppose we have $\nu$ identical copies of such a qubit. Let $\mathcal{H}_{2^{\nu}}$ denote the joint 
Hilbert space of all $\nu$ qubits. The state of $\nu$ identical copies of $|\beta_{\lambda}^{\pm}\rangle$ is given by
\begin{equation}
|\beta_{\lambda}^{\pm},\nu\rangle = \left( \cos \beta_{\lambda} |0\rangle \pm \imath \sin \beta_{\lambda}|1\rangle \right)^{\otimes \nu}\ .
\label{multiplebeta}
\end{equation}  
We now partition the Hilbert space $\mathcal{H}_{2^{\nu}}$ in two mutually complementary subspaces: $\mathcal{W}$ and $\mathcal{W}^{\perp}$ defined as
\begin{eqnarray}
\mathcal{W} &=& \bigcup_{\nu' = 1}^{(\nu/2)-1}|0\rangle^{\nu'}|1\rangle^{\nu-\nu'}\ , \nonumber \\
\mathcal{W}^{\perp} &=& \bigcup_{\nu' = \nu/2}^{\nu}|0\rangle^{\nu'}|1\rangle^{\nu - \nu'}\ . \label{Wsubspacedefine}
\end{eqnarray}
Here $|0\rangle^{\nu'}|1\rangle^{\nu - \nu'}$ denotes that subspace of $\mathcal{H}_{2^{\nu}}$ in which exactly $\nu'$ qubits are
in $|0\rangle$ state while remaining $\nu - \nu'$ qubits are in $|1\rangle$ state. 

We now find $|\langle \mathcal{W}|\beta_{\lambda}^{\pm},\nu\rangle|$, the magnitude of component of the joint qubit state $|\beta_{\lambda}^{\pm},\nu\rangle$
in the subspace $\mathcal{W}$. Its square is the probability of getting a state within $\mathcal{W}$ subspace, i.e. of getting at least half of $\nu$ 
qubits in $|1\rangle$ state after measuring $|\beta_{\lambda}^{\pm},\nu\rangle$. 
Let $\rm Prob(\nu')$ denote the probability of getting $\nu'$ qubits in $|1\rangle$ state. As the state of $\nu$ qubits is a tensor product of single qubit states, 
the probability distributions of measurement outcomes of all qubits are independent of each other. Hence $\rm Prob(\nu')$ follows the binomial distribution.
Eq. (\ref{betadefine}) implies that the probability of getting a single qubit in $|1\rangle$ state is $\sin^{2}\beta_{\lambda} = \rm Prob_{\lambda}\left(\mathcal{X}\right)$. 
Then the binomial distribution implies that $\rm Prob(\nu')$ has a sharp peak at $\nu' = \nu \times \rm Prob_{\lambda}\left(\mathcal{X}\right)$ and
decays exponentially with increasing $\nu'$ as we go away from this sharp peak. Eq. (\ref{probabilityexpressed}) implies that this sharp peak is at 
$\nu \rm Prob_{\lambda}\left(\mathcal{X}\right) = \nu\left[1-O(2^{-10})\right] \approx \nu $ for $\lambda = \lambda_{\pm}$ and $\nu \rm Prob_{\lambda}\left(\mathcal{X}\right) = \nu O(2^{-10})$ 
for $\lambda = \lambda_{\perp}$. Hence 
\begin{equation}
\lambda = \lambda_{\pm}  \Longrightarrow  \sum_{\nu' = \nu/2}^{\nu}\rm Prob (\nu') = 1- o\left(e^{-\nu}\right)
\end{equation}
and similarly $\lambda = \lambda_{\perp}$ implies that the quantity on R.H.S. is $o\left(e^{-\nu}\right)$. Here $o(x) \ll x$ is the small-$o$ notation. 
As mentioned earlier, this quantity is square of 
$|\langle \mathcal{W}|\beta_{\lambda}^{\pm},\nu\rangle|$. Hence
\begin{eqnarray}
\lambda = \lambda_{\pm} & \Longrightarrow &   |\langle \mathcal{W}|\beta_{\lambda \pm}^{\pm},\nu\rangle| = 1-o\left(e^{-\nu}\right), \nonumber \\
\lambda = \lambda_{\perp} & \Longrightarrow & |\langle \mathcal{W}|\beta_{\lambda \perp}^{\pm},\nu\rangle| = o\left(e^{-\nu}\right),
\end{eqnarray}
By choosing
\begin{equation} 
\nu = 5\ln B,
\end{equation}
we get
\begin{eqnarray}
\lambda = \lambda_{\pm} & \Longrightarrow &   |\langle \mathcal{W}|\beta_{\lambda \pm}^{\pm},\nu\rangle| = 1-o(1/B), \nonumber \\
\lambda = \lambda_{\perp} & \Longrightarrow & |\langle \mathcal{W}|\beta_{\lambda \perp}^{\pm},\nu\rangle| =  o(1/B),
\end{eqnarray}
Thus there exists a unitary operator $\mathcal{C}$ satisfying the properties of Eq. (\ref{Cdefine}) with $\epsilon = o(1/B)$ if we choose the subspaces to be
$\mathcal{Z} = \mathcal{W}$ and $\mathcal{Z}^{\perp} = \mathcal{W}^{\perp}$. If we use this operator to approximate $I_{\lambda_{\pm}}$,
we get a negligible error in our algorithm. It requires preparation of $\nu$ qubits in the state $|\beta_{\lambda}^{\pm}\rangle$ given 
by (\ref{betadefine}). The question arises: how do we prepare such a qubit? As it simulates the measurement statistics of the 
workspace state $|\phi_{\lambda}\rangle$, naturally the process of getting $|\beta_{\lambda}^{\pm}\rangle$ state has to depend upon the process of getting 
$|\phi_{\lambda}\rangle$ state which is the phase estimation algorithm. Consider the following amplitude amplification operator acting on the workspace 
\begin{equation}
\mathcal{A}\left(\phi_{\lambda},\mathcal{X}\right) = -I_{\phi_{\lambda}}I_{\mathcal{X}}
\end{equation}
where $I_{\phi_{\lambda}}$ is the selective phase inversion of $|\phi_{\lambda}\rangle$ state and $I_{\mathcal{X}}$ is the selective phase inversion of $\mathcal{X}$ subspace.
$I_{\mathcal{X}}$ is implemented using our knowledge of the subspace $\mathcal{X}$. To implement $I_{\phi_{\lambda}}$, we note that this is equivalent to the selective
phase inversion of $|\lambda\rangle|\phi_{\lambda}\rangle$ as the mainspace is in $|\lambda\rangle$ state. Now, Eq. (\ref{mainworkstate}) gives us
\begin{equation}
\mathcal{P}|\lambda,0'\rangle = |\lambda\rangle|\phi_{\lambda}\rangle,  \label{mainworkstate2}
\end{equation}
which implies that the selective phase inversion of $|\lambda\rangle|\phi_{\lambda}\rangle$ is equal to $\mathcal{P}I_{\lambda,0'}\mathcal{P}^{\dagger}$ where $I_{\lambda,0'}$
is the selective phase inversion of $|\lambda\rangle|0'\rangle$ state and it is equivalent to implementable $I_{0'}$, the selective phase inversion of $|0'\rangle$ as the 
mainspace is in $|\lambda\rangle$ state. As discussed earlier, both $\mathcal{P}$ and $\mathcal{P}^{\dagger}$ requires $2^{\mu}$ queries for implementation. Thus the operator 
$\mathcal{A}\left(\phi_{\lambda},\mathcal{X}\right)$ requires $2^{\mu +1}$ queries.

The eigenspectrum of amplitude amplification operator has been analysed in detail in Section 2 of ~\cite{qaa2}. It has two non-trivial eigenstates given by
\begin{equation}
|\mathcal{X}_{\lambda}^{\pm}\rangle = (1/\sqrt{2})\left(|\mathcal{X}_{\lambda}\rangle \pm |\mathcal{X}_{\lambda}^{\perp}\rangle\right),
\end{equation} 
where $|\mathcal{X}_{\lambda}\rangle$ ($|\mathcal{X}_{\lambda}^{\perp}\rangle$) is the projection of $|\phi_{\lambda}\rangle$ state on $\mathcal{X}$ ($\mathcal{X}^{\perp}$) 
subspace. The corresponding eigenvalues are given by 
\begin{equation}
e^{\pm\imath 2\omega_{\lambda}},\ \ \ \omega_{\lambda} = \sin^{-1}|\langle \phi_{\lambda}|\mathcal{X}_{\lambda}\rangle|\ . 
\end{equation}
Comparing above equation with Eq. (\ref{betadefine}) implies that
\begin{equation}
|\omega_{\lambda}| = |\beta_{\lambda}| \label{omegabeta}
\end{equation}
Also, it was shown in ~\cite{qaa2} that the state $|\phi_{\lambda}\rangle$ is a linear combination of eigenstates $|\mathcal{X}_{\lambda}^{\pm}\rangle$. Precisely
\begin{equation}
|\phi_{\lambda}\rangle = (1/\sqrt{2})\left(e^{\imath \omega_{\lambda}}|\mathcal{X}_{\lambda}^{+}\rangle - e^{-\imath \omega_{\lambda}}|\mathcal{X}_{\lambda}^{-}\rangle\right)
\label{linearsuper}
\end{equation}

Suppose our workspace is in an eigenstate $|\mathcal{X}_{\lambda}^{+}\rangle$ of $\mathcal{A}\left(\phi_{\lambda},\mathcal{X}\right)$ which we now denote simply by 
$\mathcal{A}$ for convenience. To get a qubit in $|\beta_{\lambda}^{\pm}\rangle$ state, we attach a qubit initially in the state $|0\rangle$ to our workspace. We apply a 
Hadamard gate on it to get the joint state of workspace and qubit as
\begin{equation}
|\mathcal{X}_{\lambda}^{+}\rangle\left(\frac{1}{\sqrt{2}}\left(|0\rangle + |1\rangle\right)\right)\ .
\end{equation}
We now apply a controlled-$\mathcal{A}$ operation which applies the operator $\mathcal{A}$ on the workspace if and only if the qubit is in $|1\rangle$ state. 
Then we get the following joint state
\begin{equation}
\frac{1}{\sqrt{2}}\left(|\mathcal{X}_{\lambda}^{+}\rangle|0\rangle + \mathcal{A}|\mathcal{X}_{\lambda}^{+}\rangle|1\rangle\right)\ .
\end{equation}
As $|\mathcal{X}_{\lambda}^{+}\rangle$ is an eigenstate of $\mathcal{A}$ with the eigenvalue $e^{\imath 2\omega_{\lambda}}$, the above state is
\begin{equation}
\frac{1}{\sqrt{2}}\left(|\mathcal{X}_{\lambda}^{+}\rangle|0\rangle + e^{\imath 2\omega_{\lambda}}|\mathcal{X}_{\lambda}^{+}\rangle|1\rangle\right)\ .
\end{equation}
which is equivalent to
\begin{equation}
|\mathcal{X}_{\lambda}^{+}\rangle\left(\frac{1}{\sqrt{2}}\left(|0\rangle + e^{\imath 2\omega_{\lambda}}|1\rangle\right)\right)\ .
\end{equation}
We now apply a Hadamard gate on the qubit and the above state becomes, upto a global phase,
\begin{equation}
|\mathcal{X}_{\lambda}^{+}\rangle\left(\cos \omega_{\lambda}|0\rangle - \imath \sin \omega_{\lambda}|1\rangle\right)\ .
\end{equation}
The qubit state is same as $|\beta_{\lambda}^{-}\rangle$ as $|\omega_{\lambda}| = |\beta_{\lambda}|$ due to Eq. (\ref{omegabeta}). 

As the workspace state remains unchanged in the above operation (it remains in the eigenstate $|\mathcal{X}_{\lambda}^{+}\rangle$ of $\mathcal{A}$), we can attach 
arbitrary, say $\nu$, number of qubits and repeat the above operation $\nu$ times using $\nu$ applications of $\mathcal{A}$ to get the state
\begin{equation}
|\mathcal{X}_{\lambda}^{+}\rangle\left(|\beta_{\lambda}^{-}\rangle\right)^{\otimes \nu}\ .
\end{equation} 
Similar arguments can show that if the workspace is in other eigenstate $|\mathcal{X}_{\lambda}^{-}\rangle$ of $\mathcal{A}$, then we get the following state
\begin{equation}
|\mathcal{X}_{\lambda}^{-}\rangle\left(|\beta_{\lambda}^{+}\rangle\right)^{\otimes \nu}\ .
\end{equation}
But in our case, the workspace is in $|\phi_{\lambda}\rangle$ state which is a linear superposition given by Eq. (\ref{linearsuper}). Hence doing above operations will
give us the state
\begin{equation}
\frac{1}{\sqrt{2}}\left(|\mathcal{X}_{\lambda}^{+}\rangle\left(|\beta_{\lambda}^{-}\rangle\right)^{\otimes \nu} + e^{\imath \tau}|\mathcal{X}_{\lambda}^{-}\rangle\left(|\beta_{\lambda}^{+}\rangle\right)^{\otimes \nu}\right)
\end{equation} 
where $\tau$ is a phase factor irrelevant for our purpose. This linear combination of $|\beta_{\lambda}^{\pm}\rangle$ states is as good as $|\beta_{\lambda}^{+}\rangle$ or
$|\beta_{\lambda}^{-}\rangle$ for our purpose as it doesn't change the measurement statistics of getting $0$'s or $1$'s.

Now we are in a position to determine the query and space complexity of the operator $\mathcal{R}$. To implement $\mathcal{R}$, we need \\
(i) One copy of $|\phi_{\lambda}\rangle$ state in the workspace. It takes $2^{\mu} = O(1/\theta_{\rm min})$ queries and $\mu = -\log_{2}\theta_{\rm min} + 16$ qubits. \\
(ii) We then need $\nu = 5\ln B$ applications of the operator $\mathcal{A}\left(\phi_{\lambda},\mathcal{X}\right)$. As each $\mathcal{A}$ takes $2^{\mu+1}$ queries, this
requires $O(\ln B 2^{\mu}) = O(\ln B/\theta_{\rm min})$ queries and $5\ln B$ qubits. \\
We need $O(B)$ applications of $\mathcal{R}$ for our algorithm to evolve $|w\rangle$ state to the target state $|t\rangle$. We add $O(B/\alpha)$ queries, required to get the
$|w\rangle$ state, to get the total query complexity of our algorithm as
\begin{equation}
O\left(\frac{B}{\alpha} + \frac{B\ln B}{\theta_{\rm min}}\right).
\end{equation}  
The number of ancilla qubits required by our algorithm is 
\begin{equation}
\mu + \nu = -\log_{2}\theta_{\rm min} + 5\ln B + 16.
\end{equation}
This is almost same as given by Eq. (\ref{mchoice1}) which uses only phase estimation algorithm to achieve the tolerable error in approximation. This concludes our discussion
on the approximation algorithm for selective phase inversion.

\section{DISCUSSION AND CONCLUSION}

We have shown in this paper that there exists a better way to transform the output state $|w\rangle$ of general quantum search algorithms to the target state $|t\rangle$ which
encodes the solution of search problems. Many important variants of the Grover's search algorithm can be shown to be special cases of the general quantum search algorithm as 
discussed in Section III of ~\cite{general}. Not just that, the general algorithm can also be used to design quantum search algorithms specifically suited for 
certain computational problems as illustrated in a recent paper~\cite{clause}. Hence we believe that the post-processing algorithm presented in this paper can find important
applications in search problems.

In Section III of this paper, we have also presented an algorithm to efficiently approximate the selective phase inversion of unknown eigenstates of any operator.
We believe that our approximation algorithm is important in its own context and can find useful applications. For example, in a recent paper~\cite{adiabaticfixed}, same 
algorithm is used in the 
context of simulating the quantum adiabatic evolution~\cite{farhiqae} on a digital quantum computer. This simulation is faster than earlier work~\cite{childs}. We can also use 
our approximation algorithm to improve the time complexity of adiabatic quantum computation with one dimensional projector Hamiltonian~\cite{adiabatic}. There, the final state of 
quantum adiabatic evolution has an overlap of $1/C$ with the desired target state, where $C$ depends upon the eigenspectrum of the initial Hamiltonian in a similar way as $B$
depends upon the eigenspectrum of $D_{s}$ in our algorithm. This final state, denoted as $|E_{-,\mu^{+}}\rangle$ in ~\cite{adiabatic}, can be transformed to the target state 
using quantum amplitude amplification provided we can implement the selective phase inversion of $|E_{-,\mu^{+}}\rangle$. Our approximation algorithm can be used for this
implementation as $|E_{-,\mu^{+}}\rangle$ is the ground state of a known Hamiltonian and it can be reliably distinguished from excited states using the phase estimation 
algorithm.       

Referring to the discussion after Eq. (\ref{X2}), it appears that we need to know $\theta_{\rm min}$ in advance to use our approximation algorithm. This knowlegde is required 
to distinguish $|\lambda_{\pm}\rangle$ eigenstates of $\mathcal{S}$ from other eigenstates by estimating their eigenvalues as $|\lambda_{\pm}| \ll \theta_{\rm min}$
and $|\lambda_{\perp}\rangle > \theta_{\rm min}$. Hence $\theta_{\rm min}$ needs to be known to do any selective transformation on $|\lambda_{\pm}\rangle$. Suppose we don't
know $\theta_{\rm min}$ and by mistake, we choose it to be such that apart from $\lambda_{\pm}$, there also exists non-$\lambda_{\pm}$ eigenphases of $\mathcal{S}$ in the 
interval $\{-\theta_{\rm min}, \theta_{\rm min}\}$. In this case, our algorithm applies a selective phase inversion on not just the $|\lambda_{\pm}\rangle$ eigenstates but also
some of the non-$|\lambda_{\pm}\rangle$ eigenstates. Similarly, if we choose $\theta_{\rm min}$ to be very small so that no eigenphase of $\mathcal{S}$ lies in the interval
$\{-\theta_{\rm min}, \theta_{\rm min}\}$ then our algorithm fails to do any selective transformation and implements a close to identity operation. In both these cases, our 
algorithm doesnot have any success gaurantee.

In case of no prior knowlegde of $\theta_{\rm min}$, we need to run multiple, say $r$, rounds of our algorithm. First, we begin by choosing $\theta_{\rm min}$ to be say 
$2^{-10}$. If it is larger than actual $\theta_{\rm min}$, then the algorithm may fail and in that case, we apply the algorithm again by choosing a decreased value 
$(1-\frac{\delta}{10})2^{-10}$. We keep on doing so and at $(r)^{\rm th}$ application of the algorithm, the chosen value of $\theta_{\rm min}$ is 
$(1-\frac{\delta}{10})^{r}2^{-10}$. For $r = O(-\ln \theta_{\rm min})$, the chosen value will be sufficiently close to the actual value $\theta_{\rm min}$ to provide a
successful quantum search algorithm.  

For the example of two-dimensional spatial search $\alpha = \theta_{\rm min} = 1/\sqrt{N}$ and $B = O(\log N)$ and hence our algorithm is just another alternative algorithm 
with a slightly increased time complexity by a factor of $O(\log \log N)$. We point out that Ambainis et.al. presented an $O(\sqrt{N\log N})$ algorithm for two-dimensional
spatial search based on classical post-processing~\cite{classicalambainis}. Basically they had shown that in case of $2d$ spatial search, w-state $|w\rangle$ is such that 
with high probability,
measuring it gives a basis state $|w'\rangle$ within the distance of $O(N^{1/4} \times N^{1/4})$ local steps from the target state $|t\rangle$. Hence, $|t\rangle$ can be 
found by classically searching a $O(N^{1/4} \times N^{1/4})$ lattice centred at $|w'\rangle$ which takes $O(N)$ time steps. Our algorithm offers an alternative based
on quantum post-processing and may be more suitable if two-dimensional spatial search is used as a quantum subroutine for other algorithms. However, we point out that our 
algorithm mainly offers advantages only for general quantum search algorithms where $\theta_{\rm min}/\log B \gg \alpha$ which is a relaxed criteria.

\end{document}